\documentclass[%
 reprint,
superscriptaddress,
showkeys,
 aps,
 pra,
]{revtex4-1}

\usepackage[margin=1in]{geometry} 
\usepackage{amsmath,amsthm,amssymb}
\usepackage{graphicx}
\usepackage{braket}
\usepackage{dsfont}
\usepackage{comment}
\usepackage{blkarray}
\usepackage{algorithm}
\usepackage[noend]{algpseudocode}
\usepackage{graphicx}% Include figure files
\usepackage{dcolumn}% Align table columns on decimal point
\usepackage{bm}% bold math
\usepackage{amsfonts}
\usepackage{epsf}  
\usepackage{appendix}
\usepackage{mathtools}
\usepackage{bm}

\usepackage{hyperref}% add hypertext capabilities
\usepackage[mathlines]{lineno}% Enable numbering of text and display math
\newtheorem{Theorem}{Theorem}

\makeatletter
\def\BState{\State\hskip-\ALG@thistlm}

\usepackage{xcolor}

\begin{document}

\preprint{APS/123-QED}

\title{Circuit implementation of bucket brigade qRAM for quantum state preparation}% Force line breaks with \\

\author{P. A. M. Casares}
 \email{pabloamo@ucm.es}
 \affiliation{Departamento de F\'isica Te\'orica, Universidad Complutense de Madrid.}%Lines break automatically or can be forced with \\
%\author{M. A. Martin-Delgado}%
 %\email{mardel@ucm.es}
%\affiliation{Departamento de F\'isica Te\'orica, Universidad Complutense de Madrid.}%



\date{\today}% It is always \today, today,

\begin{abstract}
In this short review I aim to explain how we can construct a circuit implementation of the bucket brigade qRAM first proposed in \cite{giovannetti2008quantum}. Used with classical data, this qRAM model can be used in combination with the quantum accessible data structure \cite{kerenidis2017recommendation} to prepare arbitrary quantum states quickly and repeatedly, once the data to be prepared is in memory.
\end{abstract}

\pacs{Valid PACS appear here}% PACS, the Physics and Astronomy
\keywords{qRAM, Bucket Brigade, quantum-accessible data structure.}%Use showkeys class option if keyword display desired
\maketitle

%\tableofcontents

%\section{\label{sec:intro}Introduction}
A quantum Random Access Memory is the memory structure that allows to perform the transformation
\begin{equation}
    \sum_i \alpha_i \ket{i}_{dir}\ket{0}_{dat} \rightarrow \sum_i \alpha_i \ket{i}_{dir}\ket{\psi_i}_{dat}.
\end{equation}
This memory structure plays a similar role to classical Random Access Memory, but in quantum computers, and is essential to several algorithms.

The structure of any qRAM is organised in a binary tree, where at the end of tree one finds the memory cells. The usual fanout arquitecture used most of the time in classical Random Access Memories uses the $i$-th bit in the direction register to control the switches at the $i$-th level of the tree. If there are $N= 2^n$ memory cells, this requires one bit to control $O(N)$ switches in the worst case, which is expensive in the quantum case because such entanglement is costly.

\begin{figure}
    \centering
    \includegraphics[width=200pt]{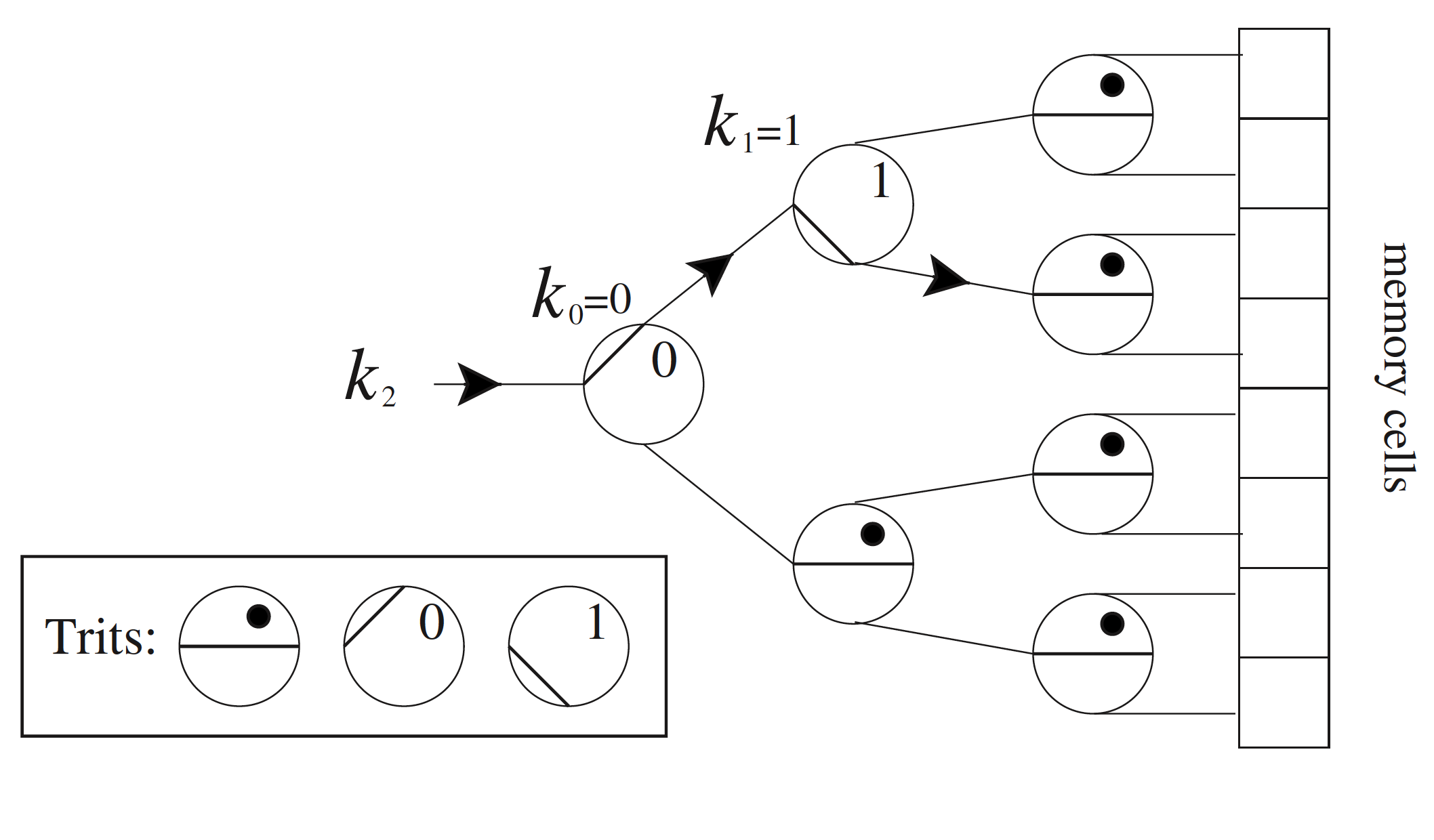}
    \caption{Scheme of the bucket brigade, taken from \cite{giovannetti2008architectures}.}
    \label{fig:bucket_brigade}
\end{figure}

Because of the previous paragraph, \cite{giovannetti2008quantum} proposed a different architecture, called "bucket brigade" which still requires $O(N)$ switches (this is unavoidable) but only activates $O(n)$ of them. This also means that much less entanglement is needed. The scheme is shown in figure \ref{fig:bucket_brigade}.

As shown in the figure, it consists on a routing algorithm which makes use of qutrits. Initially all nodes are in state $\ket{\cdot}$, except the memory cells. Suppose that the query state to the qRAM is $\ket{i}_{dir}\ket{0}_{dat}$. Then, iteratively one sends the direction bits through the tree with the following rules: if a bit $\ket{0}$ or $\ket{1}$ arrives to
\begin{itemize}
    \item a node in state $\ket{0}$, pass the signal right.
    \item a node in state $\ket{1}$, pass the signal left.
    \item a node in state $\ket{\cdot}$, save the arriving bit in the node.
\end{itemize}
After the entire direction has been sent through the tree, only $n(n-1)/2$ operations have been performed, and only $n$ qutrits have been entangled. The information can be passed via swap or C-not gates. Swap gates might be more costly, but C-not gates require to uncompute the Information Bus, that is, the qubits that move the information across the qRAM. In any case it is important to remember that the quantum state will remain entangled with the qRAM, so amplitude changes in the state will result in amplitude changes in the qRAM.

In each memory cell an arbitrary quantum state might be encoded, such that it can be quickly recovered. This however, will eliminate such state from the memory cell, since quantum states cannot be cloned. More information in the qRAMs can be found in \cite{green2019quantum}.

However, one important use of the qRAM is to work with classical data. In such case information will be encoded in the computational basis, and so can be copied. Why is this so interesting? Because, as we shall see, there is a method that leverages the use of qRAMs to prepare quantum states with real amplitudes very quickly. The theorem comes from \cite{kerenidis2017recommendation}

\begin{Theorem} \label{quantum accessible data structure}
 \cite{kerenidis2017recommendation, block-encoding_1}:
Let $M\in \mathbb{R}^{n'\times n'}$ be a matrix. If $w$ is the number of nonzero entries, there is a quantum accessible data structure of size $O(w\log^2(n'^2))$, which takes time $O(\log (n'^2))$ to store or update a single entry. 
Once the data structure is set up,  there are quantum algorithms that can perform the following maps to precision $\epsilon^{-1}$ in time $O(\text{poly}\log(n'/\epsilon))$:
\begin{equation}
    U_\mathcal{M}:\ket{i}\ket{0}\rightarrow \frac{1}{||M_{i\cdot}||}\sum_{j}M_{ij}\ket{ij};
    \label{mathcal M}
\end{equation}
\begin{equation}
    U_\mathcal{N}:\ket{0}\ket{j}\rightarrow \frac{1}{||M||_F}\sum_i ||M_{i \cdot}|| \ket{ij};
    \label{mathcal N}
\end{equation}
where $||M_{i \cdot}||$ is the $l_2-$norm of row $i$ of $M$.
This means in particular that given a vector $f$ in this data structure, we can prepare an $\epsilon$ approximation of it, $1/||v||_2 \sum_i v_i \ket{i}$, in time $O(\text{poly}\log(n'/\epsilon))$.
\end{Theorem}

Proof: 
To construct the classical data structure, create $n'$ trees, one for each row of $M$. Then, in leaf $j$ of tree $B_i$ one saves the tuple ($M_{ij}^2$, sgn($M_{ij}$)). Also, intermediate nodes are created (that join nearby branches) so that node $l$ of tree $B_i$ at depth $d$ contains the value
\begin{equation}
    B_{i,l}=\sum_{j_1,...,j_d = l}M_{ij}^2.
\end{equation}
Notice that $j_1,...,j_d$ is a string of values $0$ and $1$, as is $l$. The root node contains the value $||M_{i\cdot}||^2$.

An additional tree is created taking the root nodes of all the other trees, as the leaves of the former. One can see that the depth of the structure is polylogarithmic on $n'$, and so a single entry of $M$ can be found or updated in time polylogarithmic on $n'$.

Now, to apply $U_\mathcal{M}$, we perform the following kind of controlled rotations
\begin{equation}
\begin{split}
    &\ket{i}\ket{l}\ket{0...0}\rightarrow\\ &\ket{i}\ket{l}\frac{1}{\sqrt{B_{i,l}}}\left(\sqrt{B_{i,2l}}\ket{0}+\sqrt{B_{i,2l+1}}\ket{1}\right)\ket{0...0},
\end{split}
\end{equation}
except for the last rotation, where the sign of the leaf is included in the coefficients. It is simple to see that $U_\mathcal{N}$ is the same algorithm applied with the last tree, the one that contains $||M_{i\cdot}||$ for each $i$. Finally, for a vector, we have just one tree, and the procedure is the same. \qed

Let us see an example: suppose we want to encode the following quantum state:
\begin{equation}
    \frac{1}{\sqrt{10}}\left(-2\ket{000} + 2\ket{010} + \ket{110} - \ket{111} \right)
    \label{quantum_state}
\end{equation}

\begin{figure}
    \centering
    \includegraphics[width=200pt]{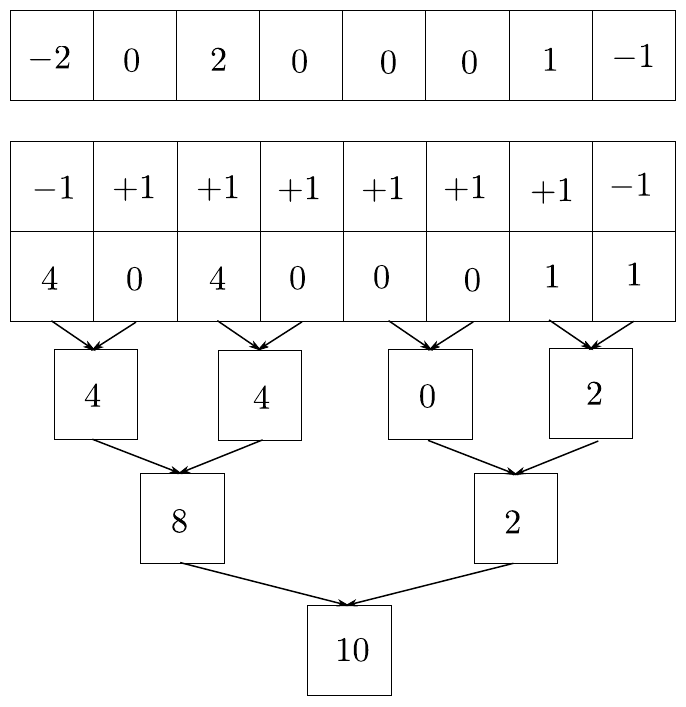}
    \caption{Quantum-accessible data structure for \eqref{quantum_state}. We need a qRAM for the first level, with memory cells $(8,2)$; another for the second level, $(4,4,0,2)$; a third one for the third level, $(4,0,4,0,0,0,1,1)$; and finally one for the signs $(1,0,0,0,0,0,0,1)$.}
    \label{fig:data_structure}
\end{figure}

As prescribed by the theorem, we would construct a data structure of figure \ref{fig:data_structure}.

The procedure that the theorem indicates is the following:
\begin{equation}
    \begin{split}
        &\ket{000} \rightarrow\\
        & \left(\sqrt{\frac{8}{10}}\ket{0} + \sqrt{\frac{2}{10}} \ket{1}\right) \otimes \ket{00} \rightarrow\\
        &   \left(\sqrt{\frac{8}{10}}\ket{0}
        \left[\sqrt{\frac{4}{8}}\ket{0} + \sqrt{\frac{4}{8}}\ket{1} \right] + \sqrt{\frac{2}{10}} \ket{1} \ket{1}\right) \otimes \ket{0}\\
        &= \left(\sqrt{\frac{4}{10}} \ket{00} + \sqrt{\frac{4}{10}} \ket{01} + \sqrt{\frac{2}{10}} \ket{11}\right)\otimes \ket{0}\rightarrow\\
        & - \sqrt{\frac{4}{10}} \ket{000} + \sqrt{\frac{4}{10}} \ket{010}\\
        &+ \sqrt{\frac{2}{10}} \ket{11}\left(\sqrt{\frac{1}{2}}\ket{0}+\sqrt{\frac{1}{2}}\ket{1} \right)\\
        &=    \frac{1}{\sqrt{10}}\left(-2\ket{000} + 2\ket{010} + \ket{110} - \ket{111} \right)
    \end{split}
\end{equation}
which is effectively what we were looking for. However notice that in order to perform the rotations quickly we need to access the quantum accessible database in superposition. 

Let us see it in the last rotation of the previous example. If before the rotation we have state 
\begin{equation}
    \sqrt{\frac{4}{10}} \ket{00} + \sqrt{\frac{4}{10}} \ket{01} + \sqrt{\frac{2}{10}} \ket{11},
\end{equation}
then we need to query how much we want to rotate the following qubit, for each of the current components. That is, we need to prepare
\begin{equation}
\begin{split}
    &\left(\sqrt{\frac{4}{10}} \ket{00}\ket{2}_a\ket{4}_b + \sqrt{\frac{4}{10}} \ket{01}\ket{2}_a\ket{4}_b\right.\\
    &\left.+ \sqrt{\frac{2}{10}} \ket{11}\ket{1}_a\ket{2}_b\right)\otimes \ket{0},
\end{split}
\end{equation}
where we want the amplitude of last qubit in state $\ket{0}$ to be $\sqrt{\frac{a}{b}}$. To do this quickly we need to query a qRAM for both $\ket{\cdot}_a$ and $\ket{\cdot}_b$. So, each row of the quantum accessible data structure must be saved in a qRAM.

\begin{figure}
    \centering
    \includegraphics[width=200pt]{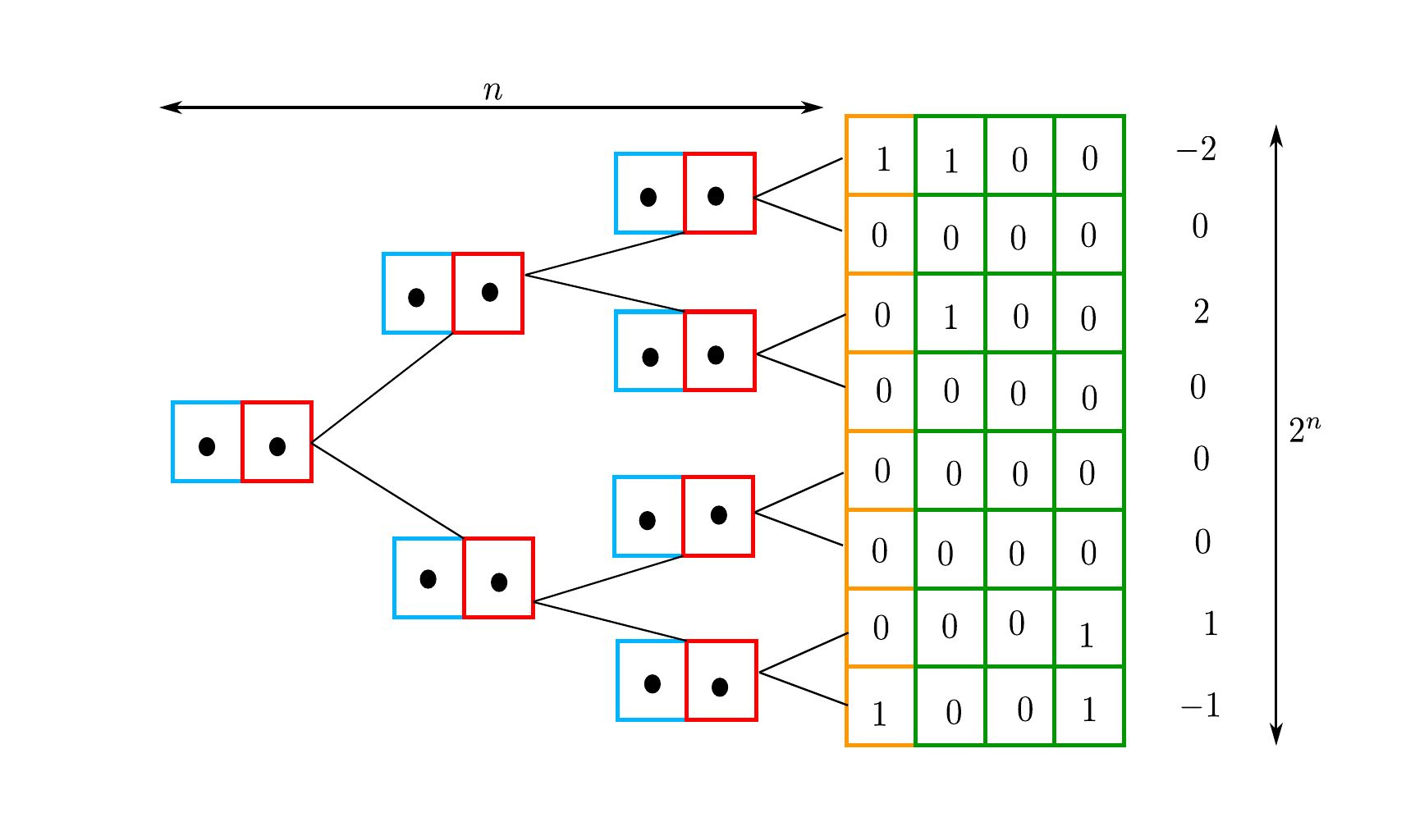}
    \caption{qRAM structure for the last row of the data structure. The blue squares indicate the switches, that can be in state $\ket{\cdot}$, $\ket{0}$ and $\ket{1}$. The red square stands for the data Bus of the qRAM structure. The orange square means the sign ($1$ for $-$) and green to encode the square of the amplitude we want to prepare}
    \label{fig:qram_initial_tree}
\end{figure}

\begin{widetext}
\begin{figure*}
    \centering
    \includegraphics[width=580pt, angle=90]{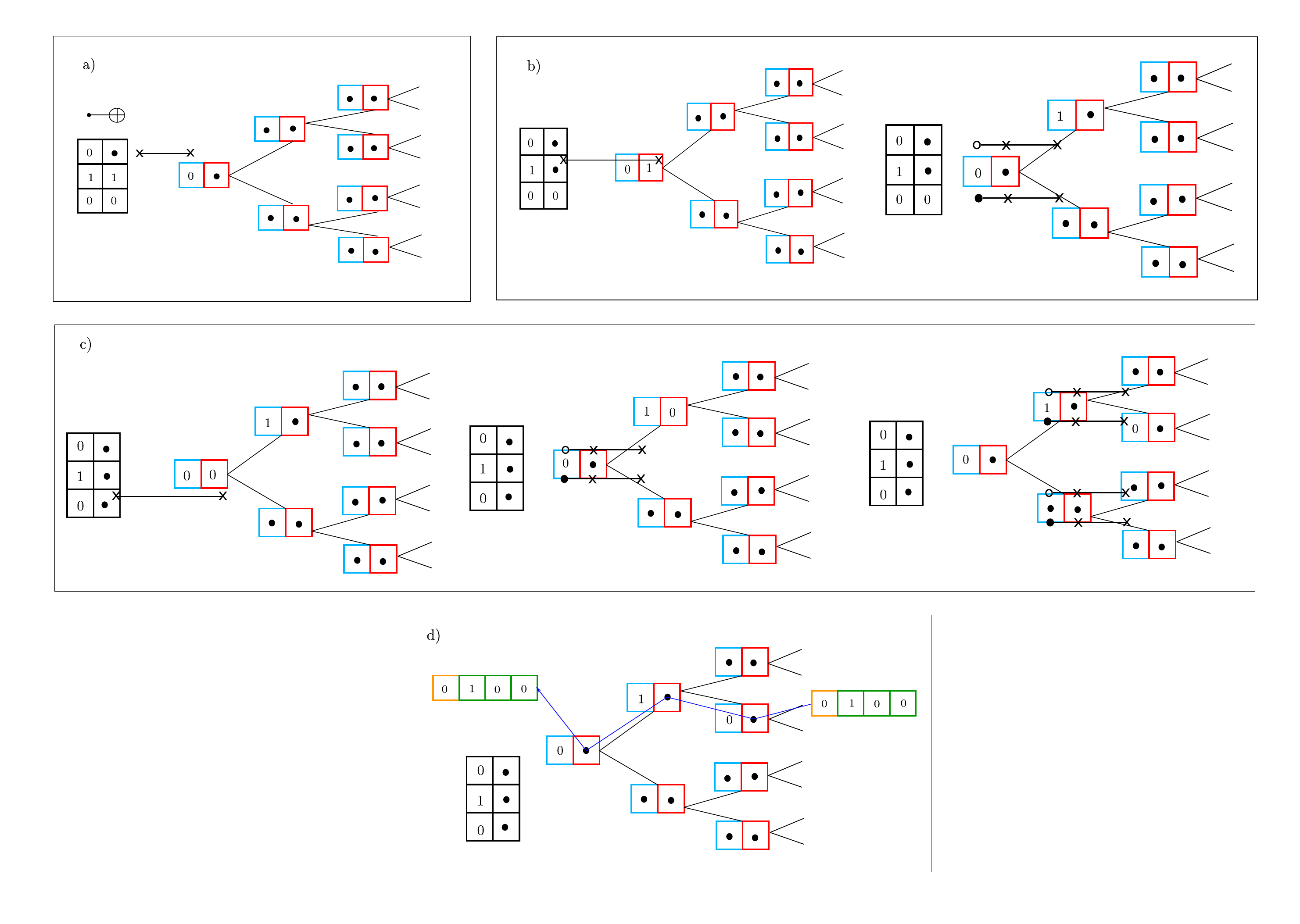}
    \caption{Example of how does the qRAM load a value. For easiness we have copied the direction register twice even though it is not necessary, the amplitudes will not be changed during the process. a) The most significant bit of the direction is saved in the first switch bit. b) The second bit of the direction is loaded in the data bus, and then directed using the first direction bit, either upwards or downwards. c) The third bit follows the same procedure: it is loaded in the data bus, and directed using the previous bit directions. d) The qRAM extracts the data.}
    \label{fig:qram_operation}
\end{figure*}
\end{widetext}

Let us now see how we would do it. We have first to prepare the qRAM of figure \ref{fig:qram_initial_tree}.
Then, we have to load both the registers $a$ and $b$ the operation shown in figure \ref{fig:qram_operation}. Notice that the operations at each level can be parallelized and performed in 2 time steps, one for the gate controlled on $\ket{0}$, and another for the gate controlled on $\ket{1}$. That means that even if the qRAM contains $O(2^n)$ qubits and switches, its operation can be performed in $O(n)$ time.

Then, once we have prepared \begin{equation}
\begin{split}
    &\left(\sqrt{\frac{4}{10}} \ket{00}\ket{4}_a\ket{4}_b + \sqrt{\frac{4}{10}} \ket{01}\ket{4}_a\ket{4}_b\right.\\
    &\left.+ \sqrt{\frac{2}{10}} \ket{11}\ket{1}_a\ket{2}_b\right)\otimes \ket{0},
\end{split}
\end{equation}
one performs the operation
\begin{equation}
\begin{split}
    &\left(\sqrt{\frac{4}{10}} \ket{000}\ket{4}_a\ket{4}_b + \sqrt{\frac{4}{10}} \ket{010}\ket{4}_a\ket{4}_b\right.\\
    &\left.+ \left(\sqrt{\frac{1}{10}} \ket{110}+ \sqrt{\frac{1}{10}} \ket{111}\right) \ket{1}_a\ket{2}_b\right)\otimes \ket{0},
\end{split}
\end{equation}
and finally uncomputes the registries $a$ and $b$ using the qRAMs again. That is
\begin{itemize}
    \item Prepare in register $b$ the state corresponding to the direction given by the previous $l-1$ bits if we are performing the $l$-th rotation. In our case, we have to query the qRAM with values $(4,4,0,2)$.
    \item To uncompute register $a$ do the same with the $l$-th level qRAM (in our case $(4,0,4,0,0,0,1,1)$, inputing the first $l-1$ bits of the state as direction bits, and a $\ket{0}$ as the least significant bit (eg $(4,4,0,1)$).
\end{itemize}
Notice that during the rotations, amplitude of ancilla registries $a$ and $b$ has not been modified, so the qRAMs will be back to their original state. 

Finally, we use the sign qRAM to perform a controlled $\pi$-phase rotation
\begin{equation}
    \begin{split}
        \frac{1}{\sqrt{10}}\left(2\ket{000} + 2\ket{010}\right.\\
        \left.+ \ket{110} + \ket{111} \right)\ket{0}_c \rightarrow\\
        \frac{1}{\sqrt{10}}\left(2\ket{000}\ket{1}_c + 2\ket{010}\ket{0}_c\right.\\
        \left. + \ket{110}\ket{0}_c + \ket{111}\ket{1}_c \right)\rightarrow\\
        \frac{1}{\sqrt{10}}\left(-2\ket{000}\ket{1}_c + 2\ket{010}\ket{0}_c \right.\\
        \left.+ \ket{110}\ket{0}_c - \ket{111}\ket{1}_c \right)\rightarrow\\
        \frac{1}{\sqrt{10}}\left(-2\ket{000}\ket{1\oplus 1}_c + 2\ket{010}\ket{0}_c\right.\\
        \left. + \ket{110}\ket{0}_c - \ket{111}\ket{1 \oplus 1}_c \right)\\
         = \frac{1}{\sqrt{10}}\left(-2\ket{000}\ket{1}_c + 2\ket{010}\ket{0}_c\right.\\
        \left.+ \ket{110}\ket{0}_c - \ket{111}\ket{1}_c \right)\\
         =\frac{1}{\sqrt{10}}\left(-2\ket{000} + 2\ket{010} + \ket{110} - \ket{111} \right)\ket{0}.
    \end{split}
\end{equation}
In fact, for the sign this might be overworking, since once the input is entangled with the qRAM one may use a $Z$ gate to perform the phase before disentangling. This technique is called a phase quickback.

So, overall we can see that the qRAMs help us prepare a quantum state without erasing the information in the qRAM in the process. This is very interesting because we can prepare the qRAMs upfront (cost $O(2^n)$) and then use them as much as one wishes at low cost each time, $O(n)$. 
\pagebreak

\bibliographystyle{ieeetr}
\bibliography{bibliography} 
\end{document}